\documentclass[10pt,english]{article}
\usepackage[T1]{fontenc}
\usepackage[latin1]{inputenc}
\usepackage[letterpaper]{geometry}
\usepackage{float}
\usepackage{amsmath}
\usepackage{amssymb}
\usepackage{graphicx}
\usepackage{esint}

\makeatletter

\usepackage{amscd}\usepackage{amsfonts}

\usepackage{babel}

\usepackage{babel}

\makeatother

\usepackage{babel}
\begin{document}

\title{
\textbf{Calibration and simulation of arbitrage effects in }\\
\textbf{a non-equilibrium quantum Black-Scholes model}\\
\textbf{by using semiclassical methods}
}

\author{Mauricio Contreras, Rely Pellicer, Daniel Santiagos and Marcelo Villena%
\thanks{Faculty of Engineering \& Sciences, Universidad Adolfo Ibáñez, Chile. %
}.}

\maketitle
\noindent An interacting Black-Scholes model for option pricing, where the usual constant interest rate $r$ is replaced by a stochastic time dependent rate $r(t)$ of the form $r(t)=r+f(t)\dot{W}(t)$,
accounting for market imperfections and prices non-alignment, was
developed in \cite{CPV}. The white noise amplitude $f(t)$, called
arbitrage bubble, generates a time dependent potential $U(t)$ which
changes the usual equilibrium dynamics of the traditional Black-Scholes
model. The purpose of this article is to tackle the inverse problem, that is, is it possible to extract the time dependent potential $U(t)$ and its associated bubble shape $f(t)$ from the real empirical financial data? In order to give an answer to this question, the interacting
Black-Scholes equation must be interpreted as a quantum Schrödinger
equation with hamiltonian operator $H=H_{0}+U(t)$, where
$H_{0}$ is the equilibrium Black-Scholes hamiltonian and $U(t)$
is the interaction term. If the $U(t)$ term is small enough, the
interaction potential can be thought as a perturbation, so one can compute the solution of the interacting Black-Scholes equation
in an approximate form by perturbation theory. In \cite{SEMICLASS}
by applying the semi-classical considerations, an approximate solution
of the non equilibrium Black-Scholes equation for an arbitrary bubble
shape $f(t)$ was developed. Using this semi-classical solution and
the knowledge about the mispricing of the financial data, one can
determinate an equation, which solutions permit obtain the functional form
of the potential term $U(t)$ and its associated bubble $f(t)$. 
In all the studied cases, the non equilibrium model performs a better
estimation of the real data than the usual equilibrium model. It is
expected that this new and simple methodology for calibrating and
simulating option pricing solutions in the presence of market imperfections,
could help to improve option pricing estimations.\\
\\
\\
\\ 
\\

\section{Introduction}

For almost 35 years, since the seminal articles by Black and Scholes
(1973, \cite{Black-Scholes}) and Merton (1973, \cite{Merton}), the
Black-Scholes (\textbf{B-S}) model has been widely used in financial
engineering to model the price of a derivative on equity. In analytic
terms, if $B(t)$ and $S(t)$ are the risk-free asset and underlying
stock prices, the price dynamics of the bond and the stock in this
model are given by the following equations:

\begin{equation}
\begin{array}{rcl}
dB(t) & = & rB(t)dt\\
dS(t) & = & \mu S(t)dt+\sigma S(t)dW(t)
\end{array}\label{BS}
\end{equation}
\\
 where $r$, $\mu$ and $\sigma$ are constants and $W(t)$ is a Wiener
process. In order to price the financial derivative, it is assumed
that it can be traded, so one can form a portfolio based on the derivative
and the underlying stock (no bonds are included). Considering only
non dividend paying assets and no consumption portfolios, the purchase
of a new portfolio must be financed only by selling from the current
portfolio. Here, $\pi(S,t)$ denote the option price, $\vec{h}(t)=(h_{S},h_{\pi})$
the portfolio and $\vec{P}(t)=(S,\pi)$ the price vector of shares.
Calling $V(t)$ the value of the portfolio at time $t$; the dynamic
of a self-financing portfolio with no consumption is given by
\begin{equation}
dV(t)=\vec{h}(t)\cdot d\vec{P}(t)\label{SF}
\end{equation}
In other words, in a model without exogenous incomes or withdrawals,
any change of value is due to changes in asset prices. \\

\noindent Another important assumption for deriving \textbf{B-S} equation
is that the market is efficient in the sense that is free from arbitrage
possibilities. This is equivalent with the fact that there exists
a self-financed portfolio with value process $V(t)$ satisfying the
dynamic:
\begin{equation}
dV(t)=rV(t)dt\label{AF}
\end{equation}
which means that any locally riskless portfolio has the same rate
of return than the bond.

\noindent For the classical model presented above, there exists a
well known solution for the price process of the derivative $\pi(t)$
(see, for example \cite{Bjork}). Given its simplicity, this formulation
can be described as one of the most popular standards in the profession.
\\

Today however, it is possible to find models that have relaxed almost
all of the initial assumptions of the Black-Scholes model, such as
models with transaction costs, different probability distribution
functions, stochastic volatility, imperfect information, etc; all
of which have improved the prediction capabilities of the original
B-S model. See \cite{Bjork}-\cite{Wilmott} for some complete reviews
of these extensions.\\

Some attempts to improve the predictions of the Black-Scholes
models, that take into account deviations of the equilibrium in the form
of arbitrage situations, have been developed in \cite{CPV}, \cite{Otto}, \cite{Panayides}, \cite{Fedotov-Panayides}. 
In this case, some of these models assume that the return
from the \textbf{B-S} portfolio is not equal to the constant risk-free
interest rate, but instead, the no arbitrage principle (\ref{AF})
is modified according to the equation
\begin{equation}
dV(t)=(r+\alpha(t))V(t)dt,\label{Arbitrage}
\end{equation}

\noindent where $\alpha(t)$ is a random arbitrage return. This formulation
gives great flexibility to the model, since $\alpha(t)$ can be seen
as any deviations of the traditional assumed equilibrium, and not
just as an arbitrage return. For instance, Ilinski (1999, \cite{Ilinski})
and Ilinski and Stepanenko (1999, \cite{Ilinski-Stepanenko}) assume
that $\alpha(t)$ follows an Ornstein-Uhlenbeck process. Deviation
from the non arbitrage assumption implies that investors can make
profit in excess from the risk-free interest rate. For example, if $\alpha(t)$
is greater than zero, then what one can do is: borrow from the bank,
paying interest rate $r$, invest in the risk-free rate stock portfolio
and make a profit. Alternatively, one could go short the option, delta
hedging it. \\

The object of this paper, is to study the arbitrage effects on the option prices.
This study will have two principal components: \\
 \\
 1) Calibration: one hopes to obtain a measure of the arbitrage effects
from the empirical financial data, and \\
 2) Simulation: the above measure can be used to obtain the ``improve''
option price and compare it with the usual Black-Scholes model and
the real option prices.\\
 \\
 For this, it is assumed that arbitrage can be modelled using equation
(\ref{Arbitrage}), so one will consider the \textbf{B-S} model in
(\ref{BS}) and self-financing portfolio condition in (\ref{SF})
and in what follows the following arbitrage condition is assumed:
\begin{equation}
dV(t)=rV(t)dt+f(S,t)V(t)dW(t)\label{SA}
\end{equation}
where $f=f(S,t)$ is a given deterministic function called ``arbitrage
bubble'' \cite{CPV} and $W$ is the same Wiener process in the dynamic
of the underlying stock $S$. Equation (\ref{SA}) will generate a
non equilibrium Black-Scholes model. Note that condition (\ref{SA})
can be rewritten as
\begin{equation}
dV(t)=rV(t)dt+f(S,t)V(t)dW(t)=\left(r+f(S,t)\dot{W}(t)\right)V(t)dt\label{CA}
\end{equation}
where $\dot{W}$ is a white noise. This can be interpreted as a stochastic
perturbation in the rate of return of the portfolio with amplitude
$f$: $\alpha=\alpha(S,t)=f(S,t)\dot{W}(t)$. \\

As it is well known, in a perfectly competitive market, assumed by
the original \textbf{B-S} model, the action of buyers and sellers
exploiting the arbitrage opportunity will cause the elimination of
the arbitrage in the very short run, so in our setting one will considered
implicitly the speed of market's adjustment by modelling an ``arbitrage
bubble'', which can be defined in duration and size, taking this
way into account the market clearance power. All this information
is contained in the function $f=f(S,t)$. In fact, in \cite{CPV}
was showed that, for an infinite arbitrage bubble $f$ the non equilibrium
Black-Scholes model goes to the usual Black-Scholes model, so (\ref{SA})
accounts implicitly for the market power clearance.\\ \\

In \cite{Otto}-\cite{Ilinski-Stepanenko} different generalizations of the Black-Scholes model are proposed. These models include a stochastic rate model whose dynamic is generated by a second Brownian motion independent of the asset Brownian motion. In a sense, these models are inspired by ``stochastic volatility ideas''. \\
What we are trying to do here, is to incorporate arbitrage effects, but as close as possible to the original Black-Scholes model, which has only one source of randomness (associated with the asset price S) and where the B bonus dynamics is completely deterministic.\\
The central idea is that arbitrage effect can change the portfolio returns in a random fashion, and the source of randomness must be generated by the same asset Brownian motion. It is in that sense that the term ``endogenous stochastic arbitrage'' appears in the title of paper \cite{CPV}. In that setting, the only remaining degree of freedom necessary is the amplitude of such a Brownian motion that is expressed in equation (\ref{SA}). \\
Although, equation (\ref{SA}) can be rewritten as a stochastic rate model as in equation (\ref{CA}), it is not clear if such interpretation is well defined in mathematical terms, or if even it is integrable. So, our point of view is not seeing our model as a stochastic rate model, but instead as a ``perturbed portfolio return model'', defined by equation (\ref{SA}). \\ \\

Thus, one is assuming a model-dependent arbitrage, where the arbitrage
possibilities are modelled with the same stochastic process that govern
the underlying stock. This assumption allows us to link the arbitrage
equation to the B-S original model%
\footnote{Otherwise, the arbitrage should be modelled exogenously to the \textbf{B-S}
model%
}. This assumption is reasonable from a theoretical perspective for
some kinds of arbitrages, which are inherent to the underlying asset,
and endogenous in nature to the asset in analysis. The validity of
this maintained hypothesis has been tested empirically bellow, see
for instance \cite{Ilinski2001}. \\

In \cite{CPV} analytical solutions of the non equilibrium Black-Scholes
model were found for a time dependent ``step function'' arbitrage
bubble $f$ for an option with maturity $T$:

\begin{equation}
f(t)={\displaystyle \left\{ \begin{array}{lll}
0 & 0\le t<T_{1}\\
f_{0} & T_{1}\le t\le T_{2}\\
0 & T_{2}<t\le T
\end{array}\right.}\label{step bubble}
\end{equation}
\\
 This particular shape of the bubble was motivated by an empirical
study of futures on the $S\&P500$ index between September 1997 and
June 2009. There, through the empirical analysis of the future mispricing,
one can get the shape of the arbitrage bubble, which in that case
corresponds roughly to a step function shape, as is showed in the figure
below:

\begin{figure}[H]
\centering \includegraphics[scale=0.5]{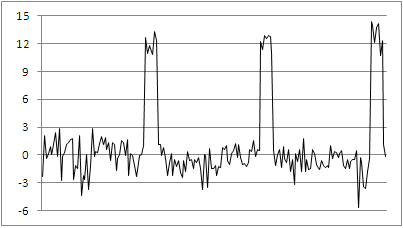} \caption{future's mispricing}
\end{figure}

So in the option pricing context, one can naturally ask: \\
 \\
 Can the shape of the arbitrage bubble $f$ be obtained from an empirical
analysis of the option mispricing, using the same approach for futures
on the $S\&P500$ index given in \cite{CPV}? \\

The object of this paper is to show that the answer is positive and to
develop a methodology for extracting the arbitrage bubble $f$ from the
empirical financial data through the analysis of the option mispricing.
In order to do that, one will need to use some result of semi-classical
approximations applied to option pricing as develop in \cite{SEMICLASS}.
There, an approximate solution for the non equilibrium Black-Scholes
equation in the presence of an arbitrary arbitrage bubble was constructed.
This semi-classical solution plus the option mispricing data, permit
to obtain a non linear equation for the arbitrage bubble. 
By solving this equation by means of numerical methods
the approximate shape of the arbitrage bubble $f$ can be obtained.
Then, taking this arbitrage bubble back to the non equilibrium Black-Scholes
equation, one can determinate the ``exact'' interacting option price
solution by means of a Crank-Nicolson method and compare it with the
usual equilibrium Black-Scholes solution. In all studied cases, the
non equilibrium solution performs a better numerical estimation for
the empirical data than the usual Black-Scholes solution. \\

To proceed and to make the paper self contained, section
2 review the interacting Black-Scholes model according to \cite{CPV} and
section 3 gives it interpretation as a quantum model. The section 4,
quickly review the principal results of applying semi-classical
quantum ideas to the interacting Black-Scholes model as developed
in \cite{SEMICLASS}. In section 5, the calibration problem is analysed,
that is, how to estimate the interaction potential in the non equilibrium
Black-Scholes framework, and the deduction of an equation which permits to
found arbitrage bubble $f(t)$ from the real financial data.
In section 6, the simulation problem is developed to obtain the exact option price solution of the non equilibrium
model, for several different data set. Finally, in section 7, final conclusion and future prospects are given.

\section{The non equilibrium Black-Scholes model}

Following \cite{CPV} we will find the price dynamics of the financial
derivative under the endogenous arbitrage condition (\ref{SA}). We
are going to derive the price dynamic as the solution $\pi(t,S)$
of certain boundary value problem. In what follows we consider the
price process depending on $t$, $S$, but we omit this dependence
for the sake of simplicity.

Using Itô calculus we get:
\[
d\pi=\frac{\partial\pi}{\partial t}dt+\frac{\partial\pi}{\partial S}dS+\frac{1}{2}\frac{\partial^{2}\pi}{\partial S^{2}}dS^{2}
\]

Given the dynamic for $S$ in (\ref{BS}) we have:

\[
d\pi=\left(\frac{\partial\pi}{\partial t}+\mu S\frac{\partial\pi}{\partial S}+\frac{\sigma^{2}}{2}S^{2}\frac{\partial^{2}\pi}{\partial S^{2}}\right)dt+\sigma S\frac{\partial\pi}{\partial S}dW
\]

Self-financing portfolio condition in (\ref{SF}) can be understood
as $dV=h_{S}dS+h_{\pi}d\pi$. Considering this and (\ref{SA}) together
and replacing dynamics for $S$ and $\pi$ we get:

\[
h_{S}\left(\mu Sdt+\sigma SdW\right)+h_{\pi}\left[\left(\frac{\partial\pi}{\partial t}+\mu S\frac{\partial\pi}{\partial S}+\frac{\sigma^{2}}{2}S^{2}\frac{\partial^{2}\pi}{\partial S^{2}}\right)dt+\sigma S\frac{\partial\pi}{\partial S}dW\right]=
\]

\[
=r\left(h_{S}S+h_{\pi}\pi\right)dt+f\left(h_{S}S+h_{\pi}\pi\right)dW
\]

Collecting $dt$- and $dW$- terms we have:

\begin{equation}
\begin{array}{rcl}
h_{S}S(\mu-r)+h_{\pi}\left({\displaystyle \frac{\partial\pi}{\partial t}+\mu S\frac{\partial\pi}{\partial S}+\frac{\sigma^{2}}{2}S^{2}\frac{\partial^{2}\pi}{\partial S^{2}}-r\pi}\right) & = & 0\\
\\
h_{S}S(\sigma-f)+h_{\pi}\left(\sigma S{\displaystyle \frac{\partial\pi}{\partial S}-f\pi}\right) & = & 0
\end{array}\label{eqn}
\end{equation}

The condition for existence of non-trivial portfolios $(h_{S},h_{\pi})$
satisfying (\ref{eqn}) give us the following:

Given the \textbf{B-S} model for a financial
market in (\ref{BS}), self-financing portfolio condition (\ref{SF})
and stochastic arbitrage condition in (\ref{SA}) the price process
$\pi$ of the derivative is the solution of the following boundary
value problem in the domain $[0,T]\times\mathbb{R}_{+}$.
\begin{equation} 
\begin{array}{rcl}
{\displaystyle \frac{\partial\pi}{\partial t}+\frac{\sigma^{2}}{2}S^{2}\frac{\partial^{2}\pi}{\partial S^{2}}+\frac{r\sigma-\mu f}{\sigma-f}\left(S\frac{\partial\pi}{\partial S}-\pi\right)} & = & 0\\
\\
\pi(T,s) & = & \Phi(s)
\end{array}\label{Main Eq}
\end{equation}
\noindent for constant $r$, $\mu$, $\sigma$, any function $f$
and a simple contingent claim $\Phi$. \\ \\ 
Thus, equation (\ref{Main Eq}) shows a particular type of arbitrage,
that occurs when the underlying asset and its arbitrage possibilities
are generated by a common and endogenous stochastic process. This
formulation is fairly general, in the sense that $f$ could take
any functional form. This function $f$ will be called the arbitrage
bubble. Note that when $f = 0$, the standard equilibrium \textbf{B-S}
model is recovered. \\ 
It is important to stress here that the model generated by equation (\ref{Main Eq}) is an out-of- equilibrium model, in the sense that, it does not satisfy the martingale hypothesis for $f\ne 0$.

\section{The interacting Black-Scholes model as a Schrödinger \protect \\
 quantum equation.}

Black-Scholes equation in the presence of an arbitrage bubble (\ref{Main Eq})
can be written as

\begin{equation} 
\check{L}_{BS}\pi+\frac{(r-\mu)f(S,t)}{\sigma-f(S,t)}\left(S\frac{\partial\pi}{\partial S}-\pi\right)=0\label{operatorBSI}
\end{equation}

\noindent where

\[
\check{L}_{BS}\pi=\frac{\partial\pi}{\partial t}+\frac{\sigma^{2}}{2}S^{2}\frac{\partial^{2}\pi}{\partial S^{2}}+r\left(S\frac{\partial\pi}{\partial S}-\pi\right)
\]

\noindent is the usual arbitrage free Black-Scholes operator. The
factor
\begin{equation} \label{Upot}
U(S,t) \equiv \frac{(r-\mu)f(S,t)}{\sigma-f(S,t)}
\end{equation}
can be interpreted as an effective potential induced by the arbitrage
bubble $f(S,t)$. In this way, the presence of arbitrage generates
an external time dependent force, which have an associated potential
$U(S,t)$. Then the interacting Black-Scholes model developed in \cite{CPV}
corresponds, from a physics point of view, to an interacting particle
with an external field force. Obviously, when arbitrage disappear,
the external potential is zero and we recover the usual Black-Scholes
dynamics. One can also see, that the option price dynamics $\pi(S,t)$
depends explicitly on the arbitrage bubble form $f(S,t)$. From a
financial optics, the arbitrage bubbles should be time-finite lapse
and they should have a characteristic amplitude. So, in general, arbitrage
bubbles can be defined by three parameters: the born-time, dead-time
and the maximum amplitude between these two times. In \cite{SEMICLASS}
an approximate analytical solution for the non equilibrium Black-Scholes
equation, for an arbitrary arbitrage bubble form was found.

\subsection{The quantum hamiltonian.}

Following \cite{SEMICLASS}, where a Black-Scholes-Schrödinger model
based on the endogenous arbitrage option pricing formulation introduced
by \cite{CPV} was developed, consider again the interacting Black-Scholes
equation (\ref{Main Eq}) and take the variable change $\xi=\ln S$,
to obtain
\[
\frac{\partial\pi}{\partial t}+\frac{\sigma^{2}}{2}\frac{\partial^{2}\pi}{\partial\xi^{2}}+\left(r-\frac{\sigma^{2}}{2}\right)\frac{\partial\pi}{\partial\xi}+\frac{(r-\mu)f}{\sigma-f}\left(\frac{\partial\pi}{\partial\xi}-\pi\right)=0.
\]
and if we make a second (time dependent) change of variables $x=\xi-(r-\frac{\sigma^{2}}{2})t$
we arrive to
\[
\frac{\partial\pi}{\partial t}+\frac{\sigma^{2}}{2}\frac{\partial^{2}\pi}{\partial x^{2}}-r\pi+\frac{(r-\mu)\check{f}}{\sigma-\check{f}}(\frac{\partial\pi}{\partial x}-\pi)=0
\]
where
\[
\check{f}(x,t)=f(e^{x+(r-\frac{\sigma^{2}}{2})t},t)
\]

Now we can state: Given the non equilibrium Black-Scholes
model in (\ref{Main Eq}) for the price of an option with arbitrage,
if we define
\[
\pi(x,t)=e^{-r(T-t)}\psi(x,t)
\]
the $\psi$ dynamics is given by
\[
\frac{\partial\psi(x,t)}{\partial t}+\frac{\sigma^{2}}{2}\frac{\partial^{2}\psi(x,t)}{\partial x^{2}}+u(x,t)\left(\frac{\partial\psi(x,t)}{\partial x}-\psi(x,t)\right)=0
\]
where
\begin{equation} \label{uupot}
u(x,t)=\frac{(r-\mu)\check{f}(x,t)}{\sigma-\check{f}(x,t)}
\end{equation}
is the interaction potential in the $(x,t)$ space. \\ \\
The last two equations can be interpreted as a
Schrödinger equation in imaginary time for a particle of mass $1/\sigma^{2}$
with wave function $\psi(x,t)$ in an external time dependent field
force generated by $u(x,t)$. If we write the Schrödinger as
\begin{equation}
\frac{\partial\psi(x,t)}{\partial t}=\check{H}\psi(x,t)\label{hpsi}
\end{equation}
Following the arguments developed by Baaquie in \cite{Baaquie} we
can read the hamiltonian operator as
\[
\check{H}=-\frac{\sigma^{2}}{2}\frac{\partial^{2}}{\partial x^{2}}-u(x,t)(\frac{\partial}{\partial x}-I)
\]
Since momentum operator in imaginary time is
\[
\check{P}=-\frac{\partial}{\partial x}\;,\qquad\qquad\check{P}^{2}=\frac{\partial^{2}}{\partial x^{2}}
\]
we finally arrive to the quantum hamiltonian for the interactive Black-Scholes
model as a function of the momentum operator.
\[
\check{H}=-\frac{\sigma^{2}}{2}\check{P}^{2}+u(x,t)(\check{P}+I)
\]

\subsection{The underlying classical mechanics.}

In order to obtain a semi-classical approximation for the solution
of the non equilibrium Black-Scholes model, we need develop the classical
equation of motion, that is, the Newton equations associated to the
quantum model. So, if we take the classical limit \textquotedbl{}$\hbar\longrightarrow0$\textquotedbl{}
the quantum hamiltonian becomes the classical hamiltonian function

\[
\mathcal{H}(x,P)=-\frac{\sigma^{2}}{2}P^{2}+u(x,t)(P+1)
\]
The classical hamiltonian equations
\[
\dot{x}=\frac{\partial\mathcal{H}}{\partial P}\ ,\ \ \ \ \ \dot{P}=-\frac{\partial\mathcal{H}}{\partial x}
\]
reduce in this case to
\[
\dot{x}=-\sigma^{2}P+u(x,t)
\]

\[
\dot{P}=-(r-\mu)(P+1)\frac{\sigma\frac{\partial\check{f}}{\partial x}}{(\sigma-\check{f})^{2}}
\]
The corresponding lagrangian

\[
\mathcal{L}=P\dot{x}-\mathcal{H}(x,P)
\]
becomes
\begin{equation}
\mathcal{L}=-\frac{1}{2\sigma^{2}}\left(\dot{x}-u(x,t)\right)^{2}-u(x,t) \label{lagrangian}
\end{equation}
The Euler-Lagrange equation
\[
\frac{d}{dt}(\frac{\partial\mathcal{L}}{\partial\dot{x}})-\frac{\partial\mathcal{L}}{\partial x}=0
\]
gives for this system, the following Newton equation
\[
\ddot{x}-\frac{\partial u}{\partial t}-[u(x,t)+\sigma^{2}]\frac{\partial u}{\partial x}=0
\]
We can consider here some special cases in detail.

\subsubsection*{The time-independent arbitrage model}

First, if the bubble depends only on S, that is $f=f(S)$, this imply
that
\[
\check{f}(x,t)=f(e^{x+(r-\frac{\sigma^{2}}{2})t})
\]
and we have in this case the identity
\[
\frac{\partial u}{\partial t}=(r-\frac{\sigma^{2}}{2})\frac{\partial u}{\partial x}
\]
so the Newton equation reads
\[
\ddot{x}-\frac{\partial}{\partial x}\left[\frac{u^{2}(x,t)}{2}\right]-\sigma^{2}\left(\frac{\partial u}{\partial x}\right)=(r-\frac{\sigma^{2}}{2})\frac{\partial u}{\partial x}
\]
or
\[
\ddot{x}-\frac{\partial}{\partial x}\left[u_{class}(x,t)\right]=0
\]
where
\[
u_{class}(x,t)=\frac{u^{2}(x,t)}{2}+(\frac{\sigma^{2}}{2}+r)u(x,t)
\]

\subsubsection*{The time-dependent arbitrage model}

In the second case, the arbitrage bubble depends only on time coordinate
$f(S,t)=f(t)$ so
\[
\check{f}(x,t)=f(e^{x-(r-\frac{\sigma^{2}}{2})t},t)=f(t)
\]
and
\[
u(x,t)=u(t)=\frac{(r-\mu)f(t)}{\sigma-f(t)}
\]
so
\[
\frac{\partial u}{\partial x}=0
\]
The Euler -Lagrange equation reads now
\[
\frac{d}{dt}(\dot{x}-u(t))=0
\]
that is
\[
\dot{x}=C+u(t)
\]
which can be easily integrated as
\begin{equation}
x(t)=Ct+\int\frac{(r-\mu)f(t)}{\sigma-f(t)}dt+D  \label{xcu}
\end{equation}
where $C$ and $D$ are arbitrary constants. \\ \\ \\
In that follows we consider arbitrage bubbles that are time dependent
only, that is, 
\[
f(S,t)=f(t)
\]
The reasons to do that are:\\
 \\
 (i) the model is more ``simple'' in mathematical terms and \\
 (ii) the financial data available to us are time dependent but no
$S$ dependent. \\
 \\
 In a further study we will analyse the behaviour of the interacting
Black-Scholes model for arbitrage bubbles that depends explicitly
on the underlying asset price $S$. \\
 Note that for the time dependent arbitrage bubble $f=f(t)$, the
$U(t)$ potential in (\ref{Upot}) and the $u(t)$ potential in (\ref{uupot}) are completely equivalent:
$U(t)=u(t)$.

\section{The semi-classical approximation}

Semi-classical methods have been used to find approximate solutions
of the Schrödinger equation in different areas of theoretical physics,
such as nuclear physics ~\cite{nuclear}, quantum gravity ~\cite{gravedad}, 
chemical reactions ~\cite{CHR}, quantum field theory ~\cite{qft},
path integrals ~\cite{integral de camino}. When the system has interactions,
the semi-classical approach gives an approximate solution for the wave
function of the system, while for free interaction case, semi-classical
approximation can give exact results ~\cite{aleman}. In this section,
following \cite{SEMICLASS} we develop a financial application, based
on the quantum arbitrage model of the previous section. \\

In a general setting, the solution of the Schrödinger equation (\ref{hpsi}) can be written
as
\[
\psi(x,t)=\int_{-\infty}^{\infty}G(xt|x'T)\Phi(x')dx'
\]
where $\Phi(x)$ is a specific contract (Call, Put, Binary Call...)
in the $x$ space, and $G(xt|x'T)$ is the propagator which admits
the path integral representation
\[
G(xt|x'T)=\int Dx(\tau)e^{A[x(\tau)]}
\]
where $A[x(\tau)]=\int_t^T \mathcal{L}(x(\tau),\dot{x}(\tau)) \ d\tau$ is the classical action evaluated over the path $x(\tau)$ ($t \leq \tau \leq T$) and the integral is done over all paths that connect the points $x(t)=x$ and $x(T)=x'$. If one writes $x(\tau)$ as $x(\tau) = x_{class}(\tau)+\eta(\tau)$ and expands the action around the classical path, one has
\[
A[x_{class}(\tau)+\eta(\tau)]=A[x_{class}(\tau)]+\frac{\delta A[\eta]}{\delta\eta}\eta+\frac{1}{2}\frac{\delta^{2}A[\eta]}{\delta\eta^{2}}\eta^{2}+...
\]
(where all functional derivatives are evaluated on the classical path $x_{class}(\tau)$) and integrate over all trajectories $\eta(\tau)$, the propagator becomes 
\[
G(xt|x'T)=e^{A[x_{class}(\tau)]}\int D\eta(t)e^{\frac{1}{2}\frac{\delta^{2}A[\eta]}{\delta\eta^{2}}\eta^{2}+...}
\]
If one consider contributions up to second order terms (see for example \cite{integral de camino}), the semi-classical approximation for
the propagator $G$ is given by
\[
G(xt|x'T)=\frac{e^{A[x_{class}(t)]}}{\sqrt{2\pi\sigma^{2}(T-t)}}
\]
On the other hand, the solution for the option price $\pi$ in the $x$ space is then
\[
\pi(x,t)=e^{(-r(T-t))}\psi(x,t)=\int_{-\infty}^{\infty}e^{(-r(T-t))}G(xt|x'T)\Phi(x')dx'
\]
so the propagator for the option price is, in the semi-classical approximation
\begin{equation} 
G_{SC}(xt|x'T)=e^{(-r(T-t))}\frac{e^{A[x_{class}(t)]}}{\sqrt{2\pi\sigma^{2}(T-t)}} \label{GSCLASS}
\end{equation}

In order to found the semi-classical approximation for the option price, in presence of a time dependent arbitrage bubble $f=f(t)$, we must
obtain first the classical solution (\ref{xcu}) for a time variable $\tau$ ($t\leq \tau \leq T$), with the initial condition $x(\tau=t)=x$ and final condition $x(\tau=T)=x'$. This implies that the constant $C$ in (\ref{xcu}) is given by
\[
C=\frac{x'-x}{T-t}-\frac{1}{T-t}\int_{t}^{T}u(\lambda)d\lambda
\]
so the Lagrangian (\ref{lagrangian}) evaluated over the classical path is
\[
\mathcal{L}(x(\tau),\dot{x}(\tau)) = -\frac{1}{2\sigma^{2}}C^{2}-u(\tau)
\]
and the action $A=\int_t^T \mathcal{L}(x(\tau),\dot{x}(\tau)) \ d\tau $ evaluated over the classical path becomes finally
\[
A[x_{class}]=-\frac{1}{2\sigma^{2}(T-t)}[(x'-x)-\rho(t,T)]^{2}-\rho(t,T)
\]
where
\begin{equation}
\rho(t,T)=\int_{t}^{T}u(\lambda)d\lambda=\int_{t}^{T}\frac{(r-\mu)f(\lambda)}{\sigma-f(\lambda)}d\lambda  \label{rrho}
\end{equation}
is the accumulative potential between $t$ and $T$. \\ \\
The semi-classical propagator in the $x$ space is then according to (\ref{GSCLASS})
\[
G_{SC}(xt|x'T)=\frac{e^{-r(T-t)}}{\sqrt{2\pi\sigma^{2}(T-t)}}e^{-\frac{1}{2\sigma^{2}(T-t)}[(x'-x)-\rho(t,T)]^{2}-\rho(t,T)}
\]
By using the transformation
\[
x=\ln(S)-(r-\frac{1}{2}\sigma^{2})t
\]
and the fact that $dx=dS/S$, one can now write the semi-classical propagator in the $(S,t)$ space as 
\begin{equation}
G_{SC}(St|S'T)=\frac{1}{S'}\frac{e^{-r(T-t)}}{\sqrt{2\pi\sigma^{2}(T-t)}}\frac{1}{e^{\rho}}e^{-\frac{1}{2\sigma^{2}(T-t)}[\ln({e^{\rho}S}/S')+(r-\frac{1}{2}\sigma^{2})(T-t)]^{2}} \label{GSISC}
\end{equation}
so the semi-classical solution for the option price is then given by
\begin{equation}
\pi_{SC}(S,t)=\int_{-\infty}^{\infty}G_{SC}(St|S'T)\Phi(S')dS'  \label{IBCSOL}
\end{equation}
Now, note that the Black-Scholes propagator is just the semi-classical propagator (\ref{GSISC}) evaluated at $\rho=0$
\begin{equation}
G_{BS}(St|S'T)=\frac{1}{S'}\frac{e^{-r(T-t)}}{\sqrt{2\pi\sigma^{2}(T-t)}}e^{-\frac{1}{2\sigma^{2}(T-t)}[\ln(S/S')+(r-\frac{1}{2}\sigma^{2})(T-t)]^{2}} \label{GBSPURE}
\end{equation}
so the pure Black-Scholes solution is
\begin{equation}
\pi_{BS}(S,t)= \int_{-\infty}^{\infty}G_{BS}(St|S'T)\Phi(S')dS'  \label{BSSOL}
\end{equation}
From (\ref{GSISC}) and  (\ref{GBSPURE}) one can see that both propagators are related by
\[
G_{SC}(St|S'T)=\frac{1}{e^{\rho}}G_{BS}(e^{\rho}St|S'T)
\]
and from (\ref{IBCSOL}) 
\[
\pi_{SC}(S,t)=\frac{1}{e^{\rho}}\int_{-\infty}^{\infty}G_{BS}(e^{\rho}St|S'T)\Phi(S')dS'
\]
which due to (\ref{BSSOL}), is equivalent to say  
\begin{equation}
\pi_{SC}(S,t)=\frac{1}{e^{\rho(t,T)}}\pi_{BS}(e^{\rho(t,T)}S,t)\label{semiclasical}
\end{equation}
The last equation therefore, is the semi-classical approximation for the non equilibrium Black-Scholes
solution for the option price, in presence of an arbitrary time dependent arbitrage bubble $f=f(t)$. 
Here $\pi_{BS}(S,t)$ is the arbitrage-free Black-Scholes solution for the specific option with contract $\Phi(S)$ and $\rho(t,T)$
is the accumulative potential given by (\ref{rrho}). \\ 
In this way, the function $\rho(t,T)$ renormalizes the bare arbitrage-free
Black-Scholes solution. One important fact of this last equation is
that it permits to obtain an approximation of our Black-Scholes-Schrödinger interacting
model from the classical Black-Scholes model, by means of a rescaling of the price variable, so usual computational codes can be easily
modified to obtain an approximation for the interacting model.

\section{Interaction potential and arbitrage bubble calibration}

Now finally, after a long trip on the interacting model and its semi-classical
approximation, we can tackle the main two point of this paper, that
is, the calibration and simulation problem for the arbitrage bubble
and for the option price solution of the non equilibrium Black-Scholes
model respectively. \\
 \\
 In order to solve the calibration problem, consider the empirical
time-series of the underlying asset $S_{emp}(t)$ and the real price
of the option $\pi_{emp}(t)$ in the interval $t\in[0,T]$. One can
ask for the interaction potential function
\[
U(t)=u(t)=\frac{(r-\mu)f(t)}{\sigma-f(t)}
\]
associated to a time dependent arbitrage bubble $f=f(t)$ that allows
the solution $\pi(S,t)$ of equation (\ref{operatorBSI}) when
evaluated over $S_{emp}(t)$ to fit all the time-serie of $\pi_{emp}(t)$.
\\

\noindent One way to proceed is to take a definite functional form
for the $U$ function with parameters $\vec{a}=(a_{0},a_{1},a_{2},...,a_{n})$.
In this case the solution of (\ref{CA}) becomes a function of the
vector $\pi=\pi(S,t,\vec{a})$ and then, the set of coefficients $\{a_{k}\}$
can be determined minimizing the quantity
\begin{equation}
\chi^{2}(\vec{a})=\sum_{k=1}^{N}(\pi(S_{emp}(t_{k}),t_{k},\vec{a})-\pi_{emp}(t_{k}))^{2}\label{chi2}
\end{equation}
over all sets of coefficients $\{a_{k}\}$. But it is not clear if
such a minimum exists or there exist several local minima and the
problem reduces to find the true one. Numerically this problem can
turn to be impossible to achieve. Moreover, our initial guess for
$U$ is a matter of taste, and it is not clear what the correct
initial functional form is and from which the $\chi^{2}$ minimization can start.
\\

\noindent In order to determine a guess function for the $U$ potential
we will follow a different path, based on the semi-classical approximation
and the notion of mispricing. The mispricing, denoted by $m(t)$,
is defined in \cite{LoMacKinlay2001} as the difference between the
empirical option price $\pi_{emp}(t)$ and the value of Black-Scholes
solution $\pi_{BS}(S,t)$ evaluated over the empirical underlying
asset price $S_{emp}(t)$
\begin{equation}
m(t)=\pi_{emp}(t)-\pi_{BS}(S_{emp}(t),t)\label{mispricing}
\end{equation}

\noindent Naturally, the function $m(t)$ above is known only over
a discrete time set of points. Let $U^{*}(t)$ be the exact potential
originated by the exact arbitrage bubble $f^{*}(t)$ which gives the
correct empirical option price when the solution of the interacting
Black-Scholes model (\ref{operatorBSI}) $\pi^{*}(S,t)$ is evaluated
over the empirical underlying asset price $S_{emp}(t)$
\begin{equation}
\pi_{emp}(t)=\pi^{*}(S_{emp}(t),t)
\end{equation}
the solution $\pi^{*}$ makes the value of the equation (\ref{chi2})
be exactly zero. Now suppose that $U^{*}(t)$ potential is weak ( $U^{*}\ll1$ ), in such a way that the semi-classical approximation for
the option price is valid, so we can replace the option price $\pi^{*}(S,t)$
by its semi-classical approximation (\ref{semiclasical})
\begin{equation}
\pi_{emp}(t)=\pi^{*}(S_{emp}(t),t)\simeq\pi_{SC}^{*}(S_{emp}(t),t)=\frac{1}{e^{\rho^{*}(t,T)}}\pi_{BS}(e^{\rho^{*}(t,T)}S_{emp}(t),t)
\end{equation}
where
\begin{equation}
\rho^{*}(t,T)=\int_{t}^{T}u^{*}(\lambda)d\lambda=\int_{t}^{T}U^{*}(\lambda)d\lambda=\int_{t}^{T}\frac{(r-\mu)f^{*}(\lambda)}{\sigma-f^{*}(\lambda)}d\lambda\label{Roaster}
\end{equation}
\\
 so the mispricing equation (\ref{mispricing}) becomes an equation for the arbitrage bubble $f^{*}(t)$ \\
\begin{equation} 
\pi_{BS}(S_{emp}(t),t)e^{\rho^{*}(t,T)}+m(t)e^{\rho^{*}(t,T)}-\pi_{BS}(e^{\rho^{*}(t,T)}S_{emp}(t),t)=0\label{Raphson}
\end{equation}
\\
 Equation (\ref{Raphson}) is the most important equation of this paper, because it allows us, from the knowledge
 about the empirical mispricing $m(t)$, to obtain an estimation of the interaction potential $U(t)$ and the arbitrage bubble $f(t)$ by doing the following steps: \\ \\ 
1) Given the empirical mispricing $m(t)$ in (\ref{mispricing}), the equation (\ref{Raphson}) can be solved for the function $\rho^{*}(t,T)$ by the Newton-Raphson method for each time instant. In this way, $\rho^{*}(t,T)$ is determinated in a discrete set of points. \\ \\ 
2) Then, by a nonlinear regression one can
estimate a continuous curve $\rho^{*}(t,T)$ that fits approximately this discrete set of points. \\ \\
3) From the definition of $\rho^{*}$ in equation
(\ref{Roaster}) we get
\begin{equation}
U^{*}(t)=-\frac{d\rho^{*}(t,T)}{dt}\label{alphasemi}
\end{equation}
and hence a time-dependent potential $U^{*}(t)$ can be determined in the weak limit from the time variation of the nonlinear regression for $\rho^{*}(t,T)$. \\ \\
4) From (\ref{Roaster}) one can obtain the arbitrage bubble $f^{*}$ according to
\begin{equation}
f^{*}(t)=\frac{\sigma U^{*}(t)}{r-\mu+U^{*}(t)}
\end{equation} \\ \\
This procedure solves the calibration problem mentioned above at least in the weak limit. For the strong regime ($U\gg1$) the semi-classical approximation could not longer be valid, but the functional form of the $U^{*}$
potential given by (\ref{alphasemi}) can still be a good starting point for obtaining an approximate value for the potential.

\section{Numerical results and option price simulation}

In order to test our method and solve the simulation problem for the
option price solution of the non equilibrium Black-Scholes model,
we simulate the behaviour of an European call option using the 90-days
futures of the e-mini S\&P 500 from September 1998 to June 2007. We
set the contract having the same underlying asset, opening and expiring
dates than the S\&P 500 futures. We establish the option strike price
as the underlying price at the opening date of the contract, assuming
the market is going to be flat, in such a way that the option price
is
\begin{equation}
\pi_{i}=max(F_{i}-K,0)
\end{equation}
where $\pi_{i}$ will be the empirical simulated option market price
at $i$-day, $F_{i}$ is the e-mini S\&P 500 future price and $K$
is the option strike price. As it is well known E-mini S\&P 500 options
are priced in index points up to two decimals. One E-mini S\&P 500
option can be exercised into one E-mini S\&P 500 futures contract
and since each contract has a multiplier of \$50, the option price
must also be multiplied by \$50 to get a corresponding dollar value
and every one point change in the price of the option or the underlying
futures for that matter is worth \$50 per contract.

\bigskip{}

\noindent Here, we specify the e-mini S\&P 500 futures contracts used
to simulate the option:

\begin{center}
\begin{tabular}{|c|}
\hline Table 1: e-mini $S\&P 500$ contracts \\ \hline \hline 
1) \; e-mini S\&P 500\; \; 12/03/1998 \;-\; 10/06/1998\\
2) \; e-mini S\&P 500\; \; 10/09/1998 \;-\; 09/12/1998\\
3) \; e-mini S\&P 500\; \; 10/12/1998 \;-\; 09/03/1999\\
4) \; e-mini S\&P 500\; \; 09/06/2005 \;-\; 07/09/2005\\
5) \; e-mini S\&P 500\; \; 07/09/2006 \;-\; 06/12/2006\\
6) \; e-mini S\&P 500\; \; 07/12/2006 \;-\; 07/03/2007\\
7) \; e-mini S\&P 500\; \; 08/03/2007 \;-\; 06/06/2007\\
\hline
\end{tabular}
\end{center}

\vspace{0.5 cm}
\noindent We will show our results in the case of the first contract
(e-mini S\&P 500 from 12/03/1998 to 10/06/1998). Firstly, we compute
the mispricing $m(t)$ in (\ref{mispricing}) between the simulated option price and the
Black-Scholes price (see figure 2). For this last calculation, we estimate
the standard deviation $\sigma$ of the underlying returns from the previous 90 days
and we take the three-months USA Treasury rate $r$ at the initial day
of the contract as the risk-free rate. The estimated numerical values in fact are $\sigma=0.0046$ and $r=0.00019$.

\begin{figure}[H]
\centering \includegraphics[scale=0.45]{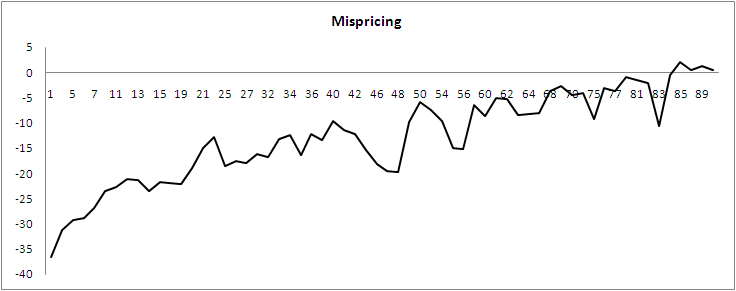} \caption{Mispricing $m(t)$}
\end{figure}
\vspace{0.5 cm}

\begin{figure}[H] 
\centering \includegraphics[scale=0.45]{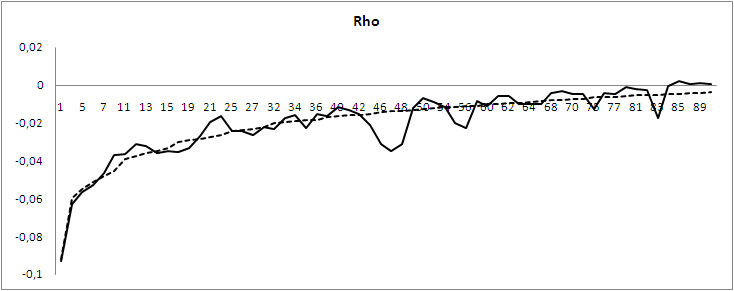} \caption{Empirical $\rho_e(t,T)$ (continuous line) and estimated $\rho(t,T)= a+bt^c$ (dashed line)}
\label{Rho}
\end{figure}

\noindent Now we can solve equation (\ref{Raphson}) via Newton-Raphson
to obtain the empirical $\rho_e(t,T)$ function daily for this contract as we can see in
Figure \ref{Rho}. We propose then a continuous potential model for this function
of the form $\rho(t,T)=a+bt^{c}$ and perform a non-linear Levenberg-Marquardt
regression in order to fit parameters $a$, $b$ and $c$. The estimated
parameter values are $a=0.1242$, $b=-0.2159$ and $c=-0.1162$ and
figure \ref{Rho} shows the results.
\vspace{0.3 cm}
\vspace{0.5 cm}

\begin{figure}[H]
\centering \includegraphics[scale=0.5]{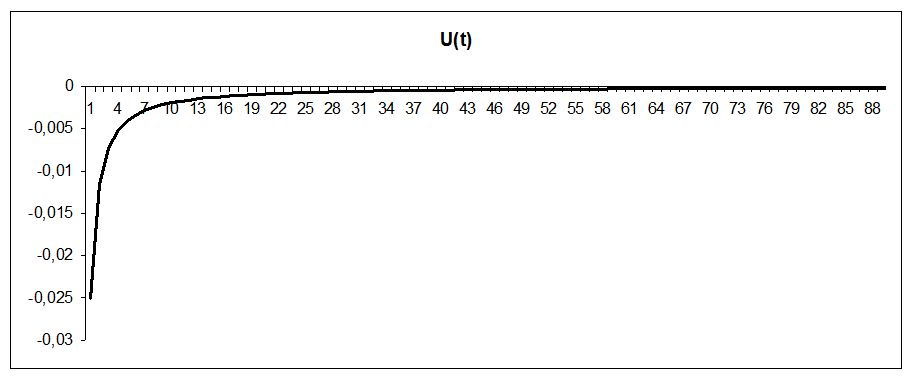}
\caption{Interacting potential $U(t)$}
\label{Alpha}
\end{figure}

\noindent At this point, we can obtain the time-dependent potential $U(t)$ by using equation (\ref{alphasemi})
\begin{equation}
U(t)=u(t)=-\frac{d\rho(t,T)}{dt}= - cbt^{c-1}
\end{equation}
as shown in figure \ref{Alpha}.
\vspace{0.3 cm}
\vspace{0.5 cm}

\begin{figure}[H]
\centering \includegraphics[scale=0.45]{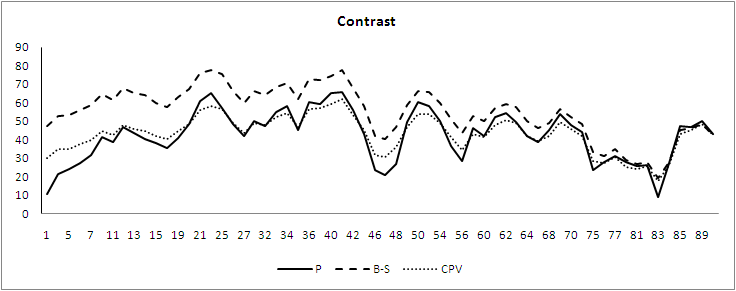}
\caption{Simulated option price P (continuous line),  Black-Scholes model price B-S (dashed line) and interacting Black-Scholes model price CPV (dotted line) for the e-mini S\&P 500 contract from 12/03/1998 to 10/06/1998.}
\label{Contrast}
\end{figure}

\noindent Now by replacing the continuous potential \ $U(t)= - cbt^{c-1}$ \ in the interacting Black-Scholes equation (\ref{operatorBSI}) and integrating it  by means of the Crank-Nicholson method, we obtain the interacting solution for the option price $\pi$ of a call option (figure \ref{Contrast}).
\vspace{0.3 cm}

\vspace{0.5 cm}

\noindent Clearly, the calibration of the potential $U(t)$ allows us to fit a more exact
price than that of the traditional Black-Scholes model without considering
arbitrage. One can test the behavior of the interacting versus
the usual Black-Scholes models for option pricing in terms of
the $\chi^{2}$ performance measure discussed before. The computed
values of the $\chi^{2}$ are: $14980.76$ for the Black-Scholes model
and $1705.44$ for the interacting Black-Scholes model, which difference is clearly visible in figure \ref{Contrast}. \\

\noindent When we use our calibrated model with its respective $U(t)$ potential for simulating
the rest of the contracts considered in our series (table 1), we find similar
results, that in all the cases defeat Black-Scholes predictions as showed in figure \ref{AllContrasts}:
\vspace{0.3 cm}
\begin{figure}[H]
\centering \includegraphics[scale=0.70]{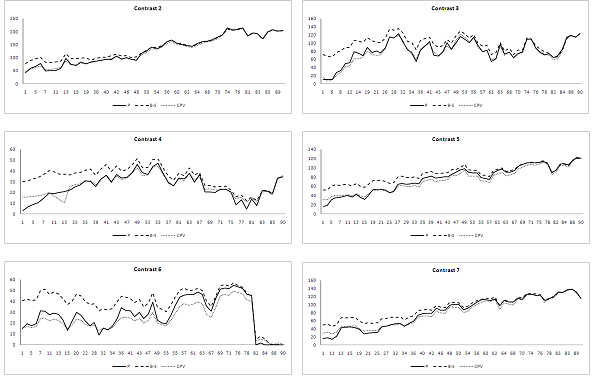}
\caption{Simulated option price P (continuous line),  Black-Scholes model price B-S (dashed line) and interacting Black-Scholes model price CPV (dotted line) for the rest of the e-mini $S\&P 500$ contracts in table 1.}
\label{AllContrasts}
\end{figure}
\vspace{0.5 cm}

\section{Conclusions}

In this work, we have calibrated the arbitrage effects for a non equilibrium
quantum Black-Scholes model of option pricing. This calibration procedure
rests heavily on the semi-classical approximation of the interacting
Black-Scholes model, which permits to construct an equation for
the interaction potential, from which the arbitrage bubble and the interaction potential can be
estimated. By using this estimated potential, we can simulate
the price trajectory of a real call option for several contracts of
the S\&P index, which allow to take into account any market imperfection
and prices desaligment. Even though we use a semi-classical approximation
for the solution of the interacting Schrödinger equation, the results
are extremely good in predicting the real option price and its trajectory
for every contract simulated.\\
 Since in real life, market imperfections always happen, almost on
a regular basis, and hence arbitrage processes form part of the normal
operation of the stock exchange, logically mispricing are always going
to exist. If we could calibrate this mispricing using the potential
of our interacting Black-Scholes, even in a small part, it is expected
that our results are always going to outperform the traditional Black-Scholes
formulation. In this context, we think this model and its calibration
procedure could be used very easily to simulate in a more exact fashion
option pricing of any underlying asset.\\
Future research could be directed to capture different potential
patterns for different underlying assets and different market situations.
Even in this case the potential is short-lived and circumstantial,
for example in the case of bubbles, rebounds, crises or critical information
(for example, when Bernanke talks!), it is possible to use our methodology
to capture the potential of the contract in a similar situation and
used to simulate the new contract. Alternatively, if the situation
is normal and no special situations are foreseen, a good practice could
be using the contract immediately before in order to calibrate the
potential and hence our quantum model; considering the reasons given
above, in almost all the cases it is expected our model will defeat
the traditional Black-Scholes model.


\begin{thebibliography}{10}
\bibitem{CPV} M. Contreras, R. Montalva, R. Pellicer and M. Villena,
\textit{Dynamic Option Pricing with Endogenous stochastic Arbitrage},
{\em Physica A: Statistical Mechanics and its Applications} Vol.
389, No. 17, (2010) 3552--3564.

\bibitem{SEMICLASS} M. Contreras, R. Pellicer, A. Ruiz and M. Villena,
\textit{A Quantum Model of Option Pricing:When Black-Scholes meets
Schrödinger and its semi-classical limit}, {\em Physica A: Statistical
Mechanics and its Applications}, Vol. 389, No. 23, (2010) 5447--5459

\bibitem{Black-Scholes} F. Black and M. Scholes, \textit{The Pricing
of Options and Corporate Liabilities.}, {\em Journal of Political
Economy.} \textbf{8,31} (1973) 637--654.

\bibitem{Merton} R.C. Merton, \textit{Theory of Rational Option Pricing.},
{\em Bell Journal of Economics and Management Science.} \textbf{4,1}
(1973) 141--183.

\bibitem{Bjork} T. Björk, \textit{Arbitrage Theory in Continuous
Time}, \newblock Oxford University Press, (1998).

\bibitem{Duffie} D. Duffie, \textit{Dynamic Asset Pricing Theory.
2nd Edition}, \newblock Princeton University Press: New Jersey,1996.

\bibitem{Hull} J. C. Hull, \textit{Options, futures, and other derivatives},
\newblock Englewood Cliffs, NJ, Prentice-Hall,1997.

\bibitem{Wilmott} P. Wilmott, \textit{Derivatives: The Theory and
Practice of financial engineering.}, \newblock J. Wiley, (1998).

\bibitem{Otto} M. Otto, \textit{Stochastic Relaxational Dynamics
Applied to Finance: Towards Non-Equilibrium Option Pricing Theory.},
{\em Internal J. Theoret. and Appl. Fin.} \textbf{3,3} (2000),
565.

\bibitem{Panayides} S. Panayides, \textit{Arbitrage opportunities
and their implications to derivative hedging.}, {\em Physica A:
Statistical Mechanics and its Applications.} Elsevier (2006).

\bibitem{Fedotov-Panayides} S. Fedotov and S. Panayides, \textit{Stochastic
arbitrage return and its implication for option pricing.}, {\em
Physica A: Statistical Mechanics and its Applications.} Elsevier
(2005).

\bibitem{Ilinski} K. Ilinski, \textit{How to Account for the Virtual
Arbitrage in the Standard Derivative Pricing.}, {\em preprint.cond-mat/9902047}
(1999).

\bibitem{Ilinski-Stepanenko} K. Ilinski and A. Stepanenko, \textit{Derivative
Pricing with Virtual Arbitrage.}, {\em preprint.cond-mat/9902046}
(1999).

\bibitem{Ilinski2001} K. Ilinski,\textit{Physics of Finance: Gauge
Modelling in Non-equilibrium pricing}, Wiley (2001).

\bibitem{Baaquie} Baaquie, B. \emph{Quantum Finance}, \newblock
Cambridge University Press, (2004).

\bibitem{nuclear} R. Schaeffer, \textit{Theoretical Methods in Medium
Energy and Heavy Ion Physics, K.W. McVoy, W.A.Friedman Eds}, \ {\em
Nato Advanced Studies Institute Series} \ B38 (1978) 109 p189.


\bibitem{gravedad} G. W. Gibbons and S. Hawking, \textit{Action integrals
and partition functions in quantum gravity}, \ {\em Phys. Rev.
D} \ vol. 15 n 10 (1977) 2752.


\bibitem{CHR} S. Keshavamurhy, \textit{semi-classical methods in
chemical reaction dynamics}, {\em Ph. D. thesis, Chemistry department
University of California}, (1994)

\bibitem{qft} V. Riva, \textit{semi-classical methods in 2D QFT:
spectra and finite size effects}, {\em Ph. D. thesis} (2004), arXiv:hep-th/0411083v1.

\bibitem{integral de camino} M. Chaichian and A. Demichev \textit{Path
Integrals in Physics Vol. I}, \ {\em Institute of Physics Publishing
IOP} (2001)

\bibitem{aleman} H. Kleinert \textit{Path Integrals in Quantum Mechanics,
Statistic, Polymer Physics, and Financial Markets.} \ {\em World
Scientific Publishing Company, 4 edition} (2006)

\bibitem{LoMacKinlay2001}Lo, A. W. and MacKinlay, A. C., \emph{A
Non-Random Walk Down Wall Street}, Princeton University Press, (1999).

\end{thebibliography}
\end{document}